\documentclass[12pt,a4paper,twoside]{article}
\usepackage{graphics,references,miscellaneous}
\footnotesep=\baselineskip
\date{\thanks{Astrophysical Dynamics 1999/2000,
              Alessandro B. Romeo (Ed.),
              Onsala Space Observatory,
              2000.}}

\begin{document}

\newcommand{\htwo}{$\mathrm{H_{2}}$}

\title{Dark Matter and Cold Fractal Clouds}

\author{Achim Tappe\\[.33cm]
        Onsala Space Observatory\\
        Chalmers University of Technology\\
        SE-43992 Onsala, Sweden\\
        (achim@oso.chalmers.se)}

\pagestyle{myheadings}

\markboth{Dark Matter and Cold Fractal Clouds}
         {Achim Tappe}

\maketitle

\begin{abstract}

There is strong evidence for a large fraction of dark matter in the Universe. Some of the evidence and
candidates for dark matter are reviewed. Dark matter in spiral galaxies may be
in the form of cold dense clouds of molecular hydrogen. This model is presented in more detail and
perspectives for detecting the cold \htwo{} are discussed.

\end{abstract}


\section{Introduction}
The question of what makes up the mass density in the Universe is of great importance to astrophysics and
cosmology. According to the widely accepted Big Bang model, the Universe originated from a very hot and dense
state and has been expanding since then. Whether this expansion will ever stop, or even reverse, depends on
the mass density (see Sect.~1.3 for more details). To determine this density it seemed at first expedient to
look at the distribution of visible matter.

The current state of research is both exciting and embarrassing. We actually do not know what makes up 99~\%
of our Universe (cf.~Sect.~1.3). Or in other words: 99~\% of stuff constituting the Universe is apparently
invisible to us (cf.~Sect.~1.2), so that we are unable to observe it directly so far. Therefore we prefer to
call this invisible matter ``dark matter''.


\subsection{What Do We Mean by ``Dark Matter''?} Let me start with some very fundamental thoughts on the role
of experiment and observation in science (see~Feynman 1998 for a more elaborate and delightful discussion on
this). Science is basically a method of finding things out. This method is based on the principle that
experiment or observation is the only judge of whether something is so or not. If one cannot answer a
question by means of experiment or observation, it is not a scientific question according to this principle.
Astronomy has a rather unique position among the sciences, since essentially all information is obtained via
observation at a distance, with no control over the experiment. One could simply say that astronomy is the
study of light (photons) that reaches Earth from space.

Now, what is then a ``dark'' object in this sense? There are basically two possibilities. The first is that
the object emits, absorbs or scatters so few photons that the photon flux, or in other words the intensity,
is too low to be detected on earth. The second is that the object is not interacting with photons at all.

To go more into detail it is useful to introduce the Standard Model of particle physics. The following
introduction is mainly based on an Internet article by Wagner (1999). The Standard model is our current
theory of elementary particles and forces. Particles whose spins are half-integer multiples of Planck's
constant $\hbar$ are called fermions (see Table 1). They make up all the matter that we see in the world
around us.
\begin{table}[!htp] 
\centering

   \begin{tabular}{|c|cc|cc|}
   \hline
   Generation&
   \multicolumn{2}{c|}{Leptons}&
   \multicolumn{2}{c|}{Quarks}\\
   \hline
   1&$\ \mathrm{e}$&$\mathrm{\nu_{e}}$&{} d&u\\
   \hdashline
   2&$\ \mathrm{\mu}$&$\mathrm{\nu_{\mu}}$&{} s&c\\
   3&$\ \mathrm{\tau}$&$\mathrm{\nu_{\tau}}$&{} b&t\\
   \hline
   \end{tabular}

\caption{Particles (fermions)}
\end{table}
\\
Important points to note are:
\begin{itemize}
   \item The fermions come in three generations. The particles in the first generation represent all the matter
   that we know about. The particles in generation 2 and 3 are almost identical to the corresponding particles
   in the first generation, except that they are more massive. They usually decay quickly into the lighter first
   generation particles.

   \item There are three charged massive leptons: the electron, the muon and the tau. Each has a neutral
   massless partner called a neutrino\footnote{Recent experiments seem to confirm that the neutrino has in fact
   a small rest mass.}.

   \item A particle that consists of quarks (named ``down'', ``up'', ``strange'', ``charm'', ``bottom'' and
   ``top'') is called a hadron. A bound state of three quarks (e.g. the proton is a `uud' state) is called a
   baryon.

   \item For each of the particles, there is a corresponding antiparticle. For example, the partner of the
   electron is the positron and the partner of the top quark is the top anti-quark. There are twelve
   particles and twelve antiparticles.
\end{itemize}
Particles whose spins are integer multiples of $\hbar$ are called bosons (see Table 2). They act as carriers
of the forces by which particles of matter interact with one other.
\begin{table}[!htp] 
\centering
   \begin{tabular}{|c|c|}
   \hline
   Force&Particle\\
   \hline
   electromagnetic&$\mathrm{\gamma}$ (photon)\\
   weak&$\mathrm{W^{{\pm}}}$, $\mathrm{Z^{0}}$\\
   strong&$\mathrm{g}$ (gluon)\\
   \hline
   \hline
   gravity&$\mathrm{G}$ (graviton)\\
   \hline
   \end{tabular}
\caption{Carriers of forces (bosons)}
\end{table}
\newpage
\noindent We note the following points:
\begin{itemize}
   \item The electromagnetic force couples in general only to charged particles (that means: light is only absorbed,
   emitted or scattered by charged particles, e.g.~neutrinos do not interact with light). However, note that the uncharged
   neutron has a small magnetic moment and therefore it does interact with light.

   \item The weak force couples to every particle, but in general its effect is only seen in the radioactive
   decay of particles.

   \item The strong force holds (``glues'') the quarks together to form the hadrons.

   \item Gravity is not really part of the Standard Model. It is described by Einstein's general theory of
   relativity. So far, there is no unification of the Standard Model, which is based upon quantum mechanics and
   general relativity. This unification is one of the most fundamental aims of modern physics (Weinberg
   1999).
\end{itemize}
Let us now come back to the original question: what does ``dark matter'' mean? We have learned the following:
\begin{itemize}
   \item[-]We use the term `matter' basically for fermions (see Table 1).

   \item[-]If we speak about `baryonic matter', we mean the normal matter made up from protons and neutrons.
   The electrons are usually included although they are not baryons but leptons (see Table 1).
   Electrons are about 2000 times lighter than protons or neutrons and are therefore often not mentioned.

   \item[-]In general, we cannot see uncharged particles, since they are not interacting with light. Neutrinos for example are
   invisible (note that invisible does not mean undetectable).
\end{itemize}
Thus dark matter could be either uncharged and therefore invisible particles or baryonic matter, whose
interaction with light is too weak to be detectable on Earth.


\subsection{Evidence for Dark Matter}
How can we detect dark matter without actually seeing it? Well, we cannot see dark matter directly, but we
can see its effect on other matter. Apart from the electromagnetic force, which acts only between charged
particles, there is the gravitational force acting between all objects with nonzero mass. As a consequence
one can infer something about the mass of an object by studying its gravitational interaction with another
object. If there is no dark matter at all, such an investigation should give us the same value for the mass
as the one inferred from observations of electromagnetic radiation.

In the following I will present two examples where this is not the case\footnote{Instructive computer
simulations can be found in the dark matter tutorial by Dursi (1998).}. These two examples provide the most
striking evidence for the presence of dark matter.

\subsubsection{The First Evidence: Clusters of Galaxies}
A galaxy cluster is  a group of a few to a few thousand galaxies, which are gravitationally bound together
but otherwise isolated in space. The Milky Way, for example, is part of the so-called Local Group with 36
counted members in total (van den Bergh 2000). It is important to note that a galaxy cluster is not a static
object. All the galaxies of the cluster have individual velocities, which can differ very much in value and
direction from the average velocity. This velocity dispersion is normally of the order of a few hundred km/s.

In the thirties, Fritz Zwicky examined the velocities of galaxies in the Coma cluster and observed a large
velocity dispersion. Assuming that the cluster is stable, i.e.~that the group is gravitationally bound and
neither collapsing nor expanding on average, he was able to infer the total mass of the cluster. In other
words, the individual velocities indicate the mass of the cluster. Galaxies with too high velocities would be
able to break free of the gravitational pull of the cluster. By assuming that the cluster is in virial
equilibrium, one can estimate the total mass (for more details see e.g. Binney \& Tremaine 1987). The
surprising result was that the visible matter was apparently not enough to explain the observed high velocity
dispersion. Much more matter would be needed to keep the cluster together. This early result was later
qualitatively confirmed by more accurate measurements. In fact, the gravitational mass of the Coma cluster
derived from images taken by the X-ray satellite ROSAT suggest that the fraction of dark matter is about
60~\% (Briel et~al. 1992).

Another possibility for determining the gravitating mass of a galaxy cluster makes use of the effect of
gravitational lensing. According to Einstein's general theory of relativity, space is curved due to the
presence of mass or energy respectively. Light is affected by this curvature in the way that it travels on a
bent path. Hence a massive object can act as a lens by means of its gravitational potential. In the case of
galaxy clusters one can observe arches or arclets, which are ``lensed'' images of background galaxies. This
is shown in the case of the galaxy cluster Abell 2218 by an impressive picture taken by the Hubble Space
Telescope (see \verb|http://oposite.stsci.edu/pubinfo/pr/2000/08/|).

The lensing effect depends on the mass of the lensing object and the spatial geometry of observer, lensing
object and lensed object. So, by knowing the geometry one can infer something about the gravitating mass of
the lensing object. This has been done in the case of Abell 2218 (Squires et~al. 1996) and the detection of
dark matter was reported.

\subsubsection{The Second Evidence: Rotation Curves of Galaxies}
For a long time it has been known that spiral galaxies spin around their center. From measuring the Doppler
shifts of stars one is able to calculate their rotational velocity $v$. To obtain a rotation curve one plots
this quantity versus the distance $R$ of the star against the galactic center. According to Newtonian dynamics
one would expect the rotation curve to fall off in proportion to $1/R^{1/2}$. Invariably, it is
observed that the stellar rotational velocity remains more or less constant with increasing distance from the
galactic center (e.g. Persic \& Salucci 1995). These facts are shown schematically in Fig.~1.
   \begin{figure}[!htp]
   \begin{center}
   \includegraphics{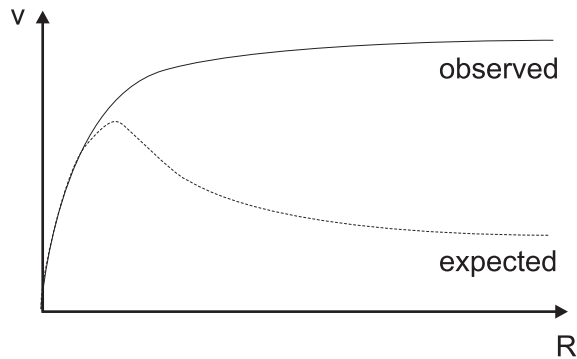}
   \caption{Expected and observed galactic rotation curves (schematic). The rotational velocity $v$ is plotted
   versus the distance $R$ from the galactic center.}
   \end{center}
   \end{figure}
If one assumes the validity of Newtonian dynamics, the observed rotation curves suggest that galaxies contain
significant amounts of dark matter. Another possible model is the so-called MOND-theory (MOdified Newtonian
Dynamics; McGaugh 1999 and references therein).


\subsection{How Much Dark Matter Is There in the Universe?}
In the introduction I stated that we do not know what makes up 99~\% of the Universe. However, some
recent reviews on cosmology claim that, for the first time, we have a plausible and complete accounting of
matter and energy in the Universe (Turner 1999, Turner \& Tyson 1999). This is not a contradiction because we
still do not have a complete understanding of all the ingredients.

In cosmology, mass or energy densities $\varrho$ respectively are expressed in terms of the so-called
critical density $\varrho \mathrm{_{crit}}$, which is necessary to give the Universe an Euclidean (flat)
geometry. The resulting dimensionless parameter is called Omega: $\Omega=\varrho/\varrho \mathrm{_{crit}}$.
Several components contribute to the total density of the Universe $\Omega\mathrm{_{total}}$. In general, one
distinguishes between:
\begin{itemize}
    \item[] $\Omega\mathrm{_{m}}$ the density of matter (note that $\Omega\mathrm{_{m}}=\Omega\mathrm{_{CDM}}+
    \Omega\mathrm{_{b}}$)

    \item[] $\Omega\mathrm{_{CDM}}$ the density of (non-baryonic) cold dark matter (see Sect.~1.4)

    \item[] $\Omega\mathrm{_{b}}$ the total density of baryonic matter (visible and dark baryonic matter)

    \item[] $\Omega\mathrm{_{vis}}$ the density of visible baryonic matter

    \item[] $\Omega_{\Lambda}$ the density of the vacuum energy (cosmological constant $\Lambda$)
\end{itemize}
Let me just briefly comment on the vacuum energy. Even a perfect vacuum has a nonzero energy according to
quantum field theory. It is called zero point or vacuum energy and can be imagined as virtual particles
coming in and out of existence. This mysterious energy behaves unusually in a way that it does not slow down
but rather speeds up the expansion of the Universe.

The prospering model of inflationary cosmology demands  $\Omega\mathrm{_{total}}=1$ (cf.~cosmology textbooks,
e.g.~Rowan-Robinson 1996 for an introduction or Coles \& Lucchin 1995 for a more advanced treatment). The
complete ``inventory'' of matter and energy totalling to $\Omega\mathrm{_{total}}=1$ is shown in Fig.~2
(adapted from Lineweaver 1999).
   \begin{figure}[!htp]
   \begin{center}
   \includegraphics{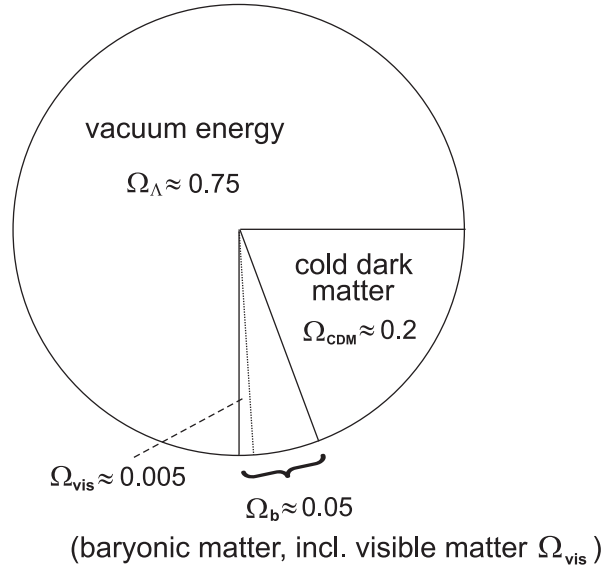}
   \caption{Matter and Energy in the Universe totalling to $\Omega\mathrm{_{total}}=1$.}
   \end{center}
   \end{figure}
As already mentioned, we do not understand all the components shown in Fig.~2. What we understand rather well
is the visible matter, which makes up only about 1~\% of all the matter and energy in the Universe. However,
intergalactic hydrogen gas, not previously visible, was recently traced with the Hubble Space Telescope
(\verb|http://oposite.stsci.edu/pubinfo/pr/2000/18/|, Tripp et al. 2000). This eventually increases
the amount of visible matter to about 2~\%.

Note here that the theory of Big Bang Nucleosynthesis in combination with measurements of the light elements
(Deuterium, Helium, Lithium) restricts the amount of baryonic matter in the Universe to
$\mathrm{\Omega_{b}\la0.05}$. A significant amount of non-baryonic matter is needed to make
$\Omega\mathrm{_{total}}=1$ and to explain the structure formation in the early Universe.

What about the implications for the fate of the Universe? Let us at first consider a cold dark matter (CDM)
model with a cosmological constant $\Lambda=0$. If the overall mass density $\Omega\mathrm{_{total}}>1$, the
expansion will reverse in the future (closed Universe). For values of $\Omega\mathrm{_{total}}<1$, the
Universe will expand forever (open Universe). In the case of $\Omega\mathrm{_{total}}=1$, the expansion of
the Universe slows down gradually until it converges to zero at infinity (flat Universe). This scenario is
shown in Fig.~3.
   \begin{figure}[!h]
   \begin{center}
   \includegraphics{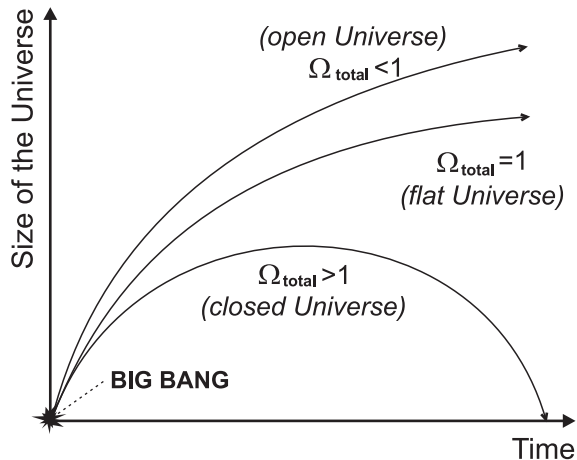}
   \caption{The parameter $\Omega\mathrm{_{total}}$ and the future of the Universe for $\Lambda=0$.}
   \end{center}
   \end{figure}
   \\
With vacuum energy present (non-zero cosmological constant), the shape of the curves in Fig.~3 and likewise the
fate of the Universe will change dramatically (see Lineweaver 1999). For example, with $\Omega\mathrm{_{total}}=1$ and
$\Omega_{\Lambda}\approx0.75$ the expansion and the size of the Universe will increase exponentially in the
future. A recent analysis of distant supernovae seems to support this hypothesis (see Perlmutter et
al.~1999).
\\
\\
However, the following questions remain (see Lineweaver 1999 and Turner 1999 for more details):
\begin{enumerate}
    \item What is the nature of the vacuum energy?

    \item What is the non-baryonic (cold) dark matter?

    \item What is the baryonic dark matter?

    \item Is $\Omega\mathrm{_{total}}$ really equal to one?
\end{enumerate}
Some advance has been made concerning the last question. A number of recent cosmic microwave background
observations show convincing evidence for a flat Universe \linebreak (e.g. BOOMERANG, Balloon Observations Of
Millimetric Extragalactic Radiation \linebreak ANd  Geomagnetics, see
\verb|http://antwrp.gsfc.nasa.gov/apod/ap000509.html| and for first results de Bernardis et al. 2000; MAT,
Microwave Anisotropy Telescope, see \linebreak \verb|http://imogen.princeton.edu/~page/matdir/www/|; Viper
telescope, Peterson et al. 2000). A recent review on vacuum energy and the cosmic background radiation is given by
Dodelson \& Knox (2000).
\\
\\
In the following section I will present some of the possible candidates for the baryonic and non-baryonic
dark matter.


\subsection{Dark Matter Candidates}
An overview of the proposed baryonic and non-baryonic dark matter candidates is given in Fig.~4.
   \begin{figure}[!htp]
   \begin{center}
   \includegraphics{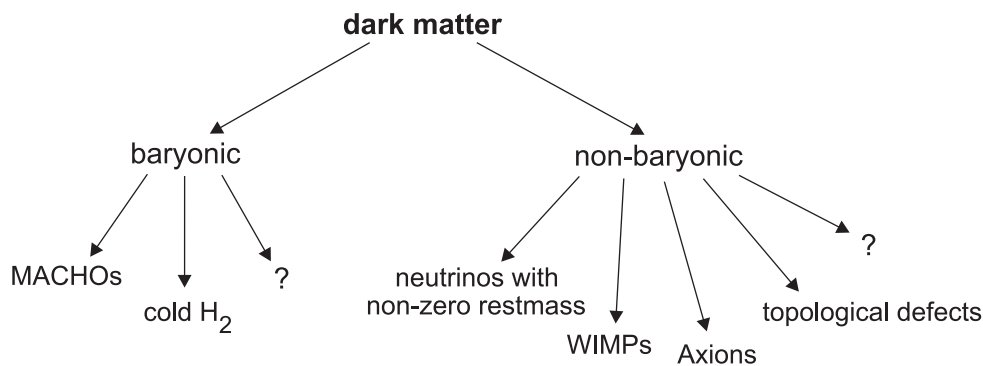}
   \caption{Candidates for baryonic and non-baryonic dark matter.}
   \end{center}
   \end{figure}
   \\
Let me briefly comment on these candidates (see Raffelt 1997 for more details). The term MACHO stands for
MAssive Compact Halo Object. Galaxies are presumably enshrouded by a halo consisting mainly of dark matter.
What is the nature of dark matter in these halos? It could be in the form of massive compact objects (e.g.
brown or white dwarfs), which are too faint to be detected from Earth. But massive objects reveal themselves
due to gravitational microlensing effects (see Sect.~1.2.1).

The search for MACHOs in the halo of the Milky Way was proposed emphatically in 1986 (see Paczynski 1986)
and began finally in 1989 (EROS, Exp\'{e}rience de Recherche d'Objects Sombres, MACHO and OGLE, Optical
Gravitational Lens Experiment, see Paczynski 1996). The first detection of a MACHO adorned the cover page of
Nature in 1993 (Alcock et al. 1993). But there are probably not enough MACHOs in
the halo to account for the observed rotation curve.

It has been proposed that dark matter in spiral galaxies may be cold molecular hydrogen distributed either in
an extended disk (Pfenniger, Combes \& Martinet 1994, hereafter PCM94, Pfenniger \& Combes 1994), or in a
spheroidal halo enshrouding the galaxy (de Paolis et al.~1995, Gerhard \& Silk 1996, de Paolis et al.~1998).
I will discuss the model of PCM94 in more detail in Sect.~2.

As for the non-baryonic dark matter, we address first the neutrinos. Having even a tiny mass, they could make
up all of the non-baryonic dark matter. Low-mass neutrinos, however, are problematic dark matter candidates
since they represent hot dark matter. The terms `cold' and `hot' dark matter refer to the early Universe.
When the Universe became transparent to radiation (about $300\,000$ years after the big bang), matter was no
longer in thermodynamic equilibrium with radiation. At that time, hot dark matter particles had relativistic
speeds, whereas the speeds of cold dark matter particles were non-relativistic. This had important
implications for the formation of structures in the early Universe. Hot dark matter basically cannot form
small-scale structures such as galaxies in the first place, which is indicated by observations. So, the idea
of low mass neutrinos accounting for most of the non-baryonic dark matter is ruled out.

What about other candidates? WIMPs (Weakly Interacting Massive particles) and axions are both non-baryonic
cold dark matter candidates. As the name implies, WIMPs are massive particles which interact or couple only
weakly with other matter. The most promising candidate is the neutralino, a particle predicted by an
extension of the Standard Model called Supersymmetry (SUSY, see Wagner 1999). The existence
of axions, which are very weakly interacting low-mass bosons, is a hot topic in theoretical physics. They
were introduced to explain the CP (Charge Parity) problem of QCD (Quantum Chromodynamics; see Raffelt 1997
for a discussion of this problem). Topological defects of space include the so-called monopoles (point-like
defects) and strings (line-shaped defects). They could have been formed during phase
transitions in the early Universe.

It is important to note that all of the non-baryonic dark matter candidates, except the neutrinos, have not
been observed yet. Their existence, and the validity of the theories predicting them, is still controversial.
Laboratory searches and astronomical observations regarding these exotic particles and structures are among
the most important scientific enterprises for understanding the Universe.


\section{Molecular Hydrogen as a Dark Matter Candidate}
Cold molecular hydrogen is a possible candidate for baryonic dark matter (see Sect.~1.4). PCM94 suggested
that a large fraction of dark matter in spiral galaxies is cold molecular hydrogen. I will discuss the
proposed model in the following section. In Sect.~2.2, I briefly comment on some perspectives for detecting
cold molecular hydrogen.

\subsection{The Cold Fractal Cloud Model}
PCM94 proposed that dark matter in spiral galaxies may be in the form of cold fractal clouds in an extended
disk. The smallest building blocks of these clouds are cold, dense clumps of molecular hydrogen gas, called
``clumpuscules''. According to the model of PCM94, the clumpuscules have a radius of the order of 30 AU,
which is about the radius of the solar system, and a mass of approximately a Jupiter, that is
$10^{-3}\,\mathrm{M_{\odot}}$. The density is about $10^{9}$ H atoms per $\mathrm{cm^{3}}$.

By discussing some fundamental questions, I would like to emphasize some of the supporting as well as critical
arguments concerning this model.

\subsubsection{Why Is Molecular Hydrogen a Good Candidate for Dark Matter in Spiral Galaxies?}
First of all, hydrogen is by far the most abundant element in the Universe. So, one might as well suspect
that molecular hydrogen (\htwo) is the most abundant molecule. It is a symmetric diatomic molecule and has no
permanent dipole moment. Therefore electromagnetic dipole transitions are forbidden. \htwo{} has, however, a
small permanent quadrupole moment. As a consequence, \htwo{} shows no strong rotational-vibrational (rovib)
spectrum. What you can observe instead are weak rovib and pure rotational lines in the infrared, due to
electromagnetic quadrupole transitions. These lines can only be seen when the molecular hydrogen is excited,
e.g. caused by thermal collisions or absorption of UV starlight followed by fluorescence. Another possibility
is to observe \htwo{} lines in absorption against IR sources, e.g. young stellar objects (YSOs). The
resulting lines are usually weak, typically only a few percent of the continuum, which makes the observations
very difficult. \htwo{} is well-studied in absorption against UV sources, usually O- or B-type stars. But the
extinction and scattering of UV radiation by interstellar dust makes these observations either difficult or
sometimes impossible, particularly in the case of dust-obscured or very distant regions.

To cut a long story short, there is probably a great deal of \htwo{} which we cannot see with current
instruments, especially when it is cold, that means below 100K, and not exposed to UV radiation. It is
therefore a good candidate for baryonic dark matter.

Another supporting argument for \htwo{} as a dark matter candidate is its consistency with the evolution of
spiral galaxies along the Hubble Sequence (see Fig.~5). It was proposed that the evolution goes from late to
early Hubble types (Pfenniger, Combes \& Martinet 1996). During that process the fraction of dark matter
seems to decrease but the amount of visible stars increases. The most straightforward explanation is that
dark matter is at least partly consumed by star formation in the visible disk of spiral galaxies. Hydrogen
would therefore be a reasonable candidate for dark matter in spiral galaxies.
   \begin{figure}[!htp]
   \begin{center}
   \includegraphics{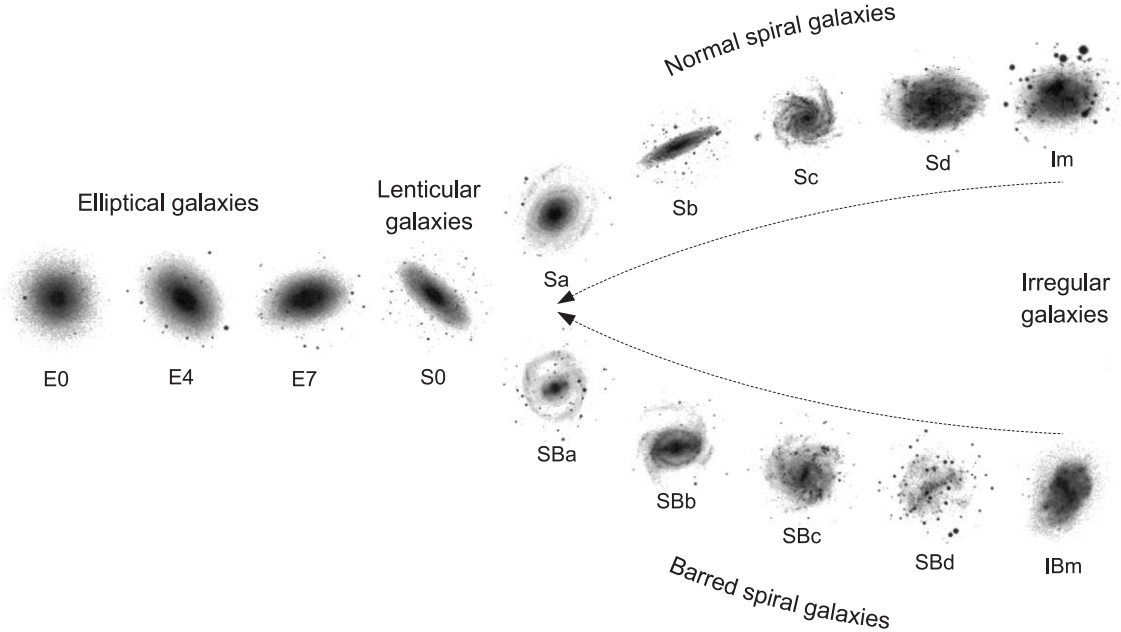}
   \caption{The Hubble Sequence of galaxies. The proposed evolution of spiral galaxies is indicated by the
   dashed arrows (Fig. adapted from Laustsen et al. 1987).}
   \end{center}
   \end{figure}
\subsubsection{Why Is the $\mathbf{H_{2}}$ Cold?}
No detailed model describing the thermal state of the dark \htwo{} clouds exists so far. This is
mainly because the precise heating rates and mechanisms in the outer regions of galaxies are unknown. So, the
thermal stability of cold clouds in an extended visible disk is not obvious. The heating by cosmic rays, for
example, can probably be an important factor governing the temperatures in steady state with the cooling in
the clouds (Walker \& Wardle  1999, Sciama 2000).

However, one can try to set approximate upper limits to the temperature. High temperatures will eventually
result in the evaporation of the denser clouds into the hot HI-gas component of the Galactic halo. This and
observational constraints probably limit the temperatures of the dense \htwo{} clouds to values of about 10K or below.
PCM94 suppose that they are in thermal equilibrium with the cosmic background radiation and hence have a
temperature of around 3K.

\subsubsection{What Determines the Size, Density and Distribution of the Clumpuscules?}
If the clumpuscules are dense ($\mathrm{n_{H} \approx 10^{9} cm^{-3}}$) and have low temperatures ($\mathrm{T
\la 10K}$) then the critical question is why the gas clouds do not form stars. Presumably any such gas
clouds have spent a long time, possibly several Gyr, in the extended disk. Why was there no star formation
even at low efficiency?

Pfenniger and Combes propose that the clouds have a fractal structure with the clumpuscules representing the
smallest building blocks (Pfenniger \& Combes 1994). This is not implausible since the fractal nature of the
interstellar cold gas is rather well known (Falgarone et al.~1992, Pfenniger 1996, Combes 1999). An analogy
of such a fractal distribution is shown in Fig.~6.
   \begin{figure}[!htp]
   \begin{center}
   \includegraphics{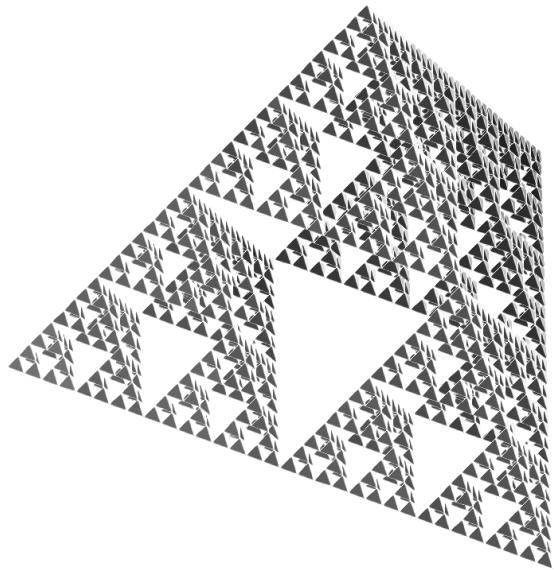}
   \caption{A fractal skewed web (Sierpinski tetrahedron, cf. Mandelbrot 1983)}
   \end{center}
   \end{figure}

A basic feature of every fractal structure is its self-similarity. Going to smaller and smaller scales one
encounters the same structure again and again (see Fig.~6). In theory this behaviour would continue to
infinitely small scales. In reality fractals show self-similarity only over a certain number of scales since
there is a natural limit for the size of the building blocks.
It is important to note that physical properties of fractal objects could be quite different from normal,
smooth objects (e.g.~objects with a roughly fractal surface have a much bigger effective surface area).

What about the implications of the fractal distribution? First of all, space would not be
occupied homogeneously. Actually there are large regions of emptiness (again see Fig.~6) making the detection
of such objects more difficult. But the clumpuscules themselves are in fact closer together, meaning that
there are local regions with high concentration. That makes the collision timescale of the clumpuscules
smaller than their collapse time. In other words, frequent collisions prevent the clumpuscules from
gravitational collapse and hence from forming stars. The proposed size and density can be understood as a
result of the dynamical equilibrium between collisions and gravitational collapse (see Pfenniger \& Combes
1994 for a more detailed discussion).

\subsection{Perspectives for Detecting Cold $\mathbf{H_{2}}$ in the Galactic Disk and Halo}
Many observational probes for detecting cold gas clouds of molecular hydrogen in an extended disk of our
Galaxy have been suggested (e.g.~Combes and Pfenniger 1997). The most promising are the study of quasar absorption
lines and lensing effects.

\subsubsection{Quasar Absorption Lines}
Absorption is a promising way to trace the cold \htwo{} gas (see Sect.~2.1.1). The absorption due to
\htwo{} in the outer parts of our Galaxy could, in principle, be detected in the UV spectra of quasars.
However, there are several difficulties.

First of all, \htwo{} absorbs in the UV at wavelengths around 1000 \AA{} (Lyman and Werner bands). Ground
based observations are impossible because of strong atmospheric absorption. Only the recently launched FUSE
satellite (Far Ultraviolet Spectroscopic Explorer, see \verb|http://fuse.pha.jhu.edu|) is currently able to
perform measurements at these wavelengths from space.

Another complication is the line damping due to the proposed high column densities
($\mathrm{N_{H}\approx10^{25}\,cm^{-2}}$). The absorption lines would become extremely saturated and
broadened (natural line widths far exceed the Doppler widths). As a consequence, all \htwo{} lines in the UV
overlap, meaning that all the light with wavelengths $\mathrm{\lambda\la3000\,\AA}$ is absorbed (see
Combes and Pfenniger 1997 for a simulated absorption profile). These absorptions may appear and disappear on
scales of one year, since the clumpuscules are supposed to have a proper motion of around 100 km
$\mathrm{s^{-1}}$.

The detection probability of such an absorption event is unclear. Due to the fractal distribution of
the clumpuscules (see Sect.~2.1.3), the surface filling factor is less than 1~\%. That would make
absorption events very rare and the chance of detection with a satellite like FUSE would be very low.

\subsubsection{Gravitational Microlensing and Gaseous Lensing}
Another promising approach for detecting dark matter in our Galaxy makes use of gravitational microlensing
effects. MACHOs have already been detected by that method (see Sect.~1.4). The principle of gravitational
lensing is shown in Fig.~7:
   \begin{figure}[!htp]
   \begin{center}
   \includegraphics{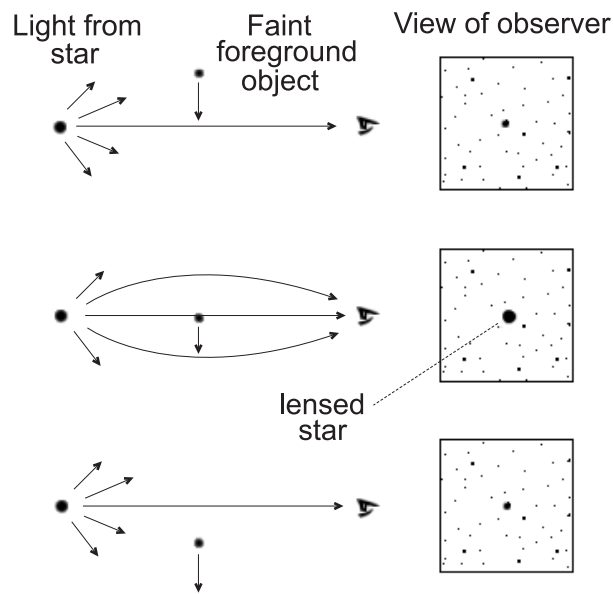}
   \caption{Gravitational lensing of a background star (schematic, adapted from Brau 1999).}
   \end{center}
   \end{figure}
   \\
The foreground object in Fig.~7 can be any massive object. The mass and proper motion of the object mainly
determine the duration of the lensing event (Paczynski 1996). The clumpuscules as proposed by PCM94 are not
able to produce measurable effects through gravitational lensing since their mass or column density is too
small. They could, however, cause lensing effects due to gaseous lensing (Draine 1998), i.e.~refraction of
optical starlight passing through the clouds. In fact, lensing events resulting from MACHOs and from
clumpuscules would resemble each other. Rafikov and Draine (2000) put constraints on cold \htwo{} clouds
from the results of existing gravitational microlensing searches. But there is a way to distinguish
between MACHO and gaseous lensing events. If the lensing object consists of \htwo{} gas, absorption lines
imposed on the stellar spectrum could be seen (Draine 1998). Supplementing any existing searches for
gravitational microlensing events by spectroscopy would allow the detection of such gas clouds.

Another method for detecting dark matter is to study the distortion of background galaxy images caused by
gravitational lensing. Large samples are needed to perform statistics and obtain information on the
distribution of dark matter. The first results of two surveys were presented recently (see Fischer et
al.~1999, \verb|http://www.sdss.org| and Van Waerbeke et al.~2000,
\verb|http://www.cfht.hawaii.edu/News/Lensing|).

\subsubsection{Extreme Scattering Events}

Extreme scattering events are significant flux changes detected whilst observing compact radio quasars
(Fiedler et al.~1994). The variations are believed to originate from refractive effects due to plasma lenses
crossing the line of sight. Photoionized material and free electrons around the clumpuscules may
explain these events (Walker and Wardle 1998).

The photoionized ``skin'' of a clumpuscule is produced by its interaction with the cosmic ray background
radiation and UV starlight. The cosmic rays would also produce $\gamma$-ray emission, which is possibly explaining the
observed $\gamma$-ray background emission (Kalberla et al. 1999, de Paolis et al. 1999).


\section{Conclusions}
Only about 1~\% of the matter in our Universe is visible to us. About 25~\% is dark matter and can only be
detected via its gravitational interaction with the visible matter. The rest is probably vacuum energy and
could make $\Omega\mathrm{_{total}}=1$ (flat Universe).

According to the theory of Big Bang Nucleosynthesis
most of the dark matter has to be non-baryonic. But there is strong evidence that spiral galaxies contain a
large amount of baryonic matter, presumably in the form of cold dense clouds of molecular hydrogen. The most
promising perspectives for detecting them are quasar absorption lines and lensing events.

However, our current models of dark matter are still rather vague and tentative. Observations and experiments
shedding light on its true nature are hard to perform. This quest leads us to the very foundations of
modern physics and cosmology and will almost certainly have a deep impact on our understanding of the
Universe.


\section{Questions and Discussion}
This is a summary of the questions asked during the discussion after the talk.

\paragraph{}
\textbf{Dr. Thomasson} asked: If you derive galaxy cluster masses you are assuming that they are in
virial equilibrium. Is that a reasonable assumption?
\paragraph{}
\textbf{A. Tappe} answered: Yes, it is a reasonable assumption if the age of the cluster is much larger than
its dynamical timescale. However, departures from virial equilibrium can be significant, especially in the
outer parts. Besides, the assumption of virial equilibrium is the only way to get a simple estimate of the
total mass of a cluster with a given velocity distribution. The errors are expected to be comparable to the
statistical uncertainties of the measurements and should be at least an order of magnitude less than the mass
estimated for the dark matter. For a more complete discussion see Binney \& Tremaine (1987, pp. 26-29 and
610-616).
\paragraph{}
\textbf{Dr. Thomasson} continued: How do spiral galaxies evolve along the Hubble Sequence from type Sd to Sa
by forming stars from \htwo{} gas in the extended disk? How is this gas able to form stars?
\paragraph{}
\textbf{A. Tappe} replied: I pointed out that spiral galaxies evolve along the Hubble Sequence from late to
early types. During that process dark matter seems to be consumed by star formation. According to the model
of PCM94 \htwo{} in spiral galaxies is in the form of cold clouds of molecular hydrogen. Frequent collisions
among the clumpuscules, the smallest building blocks of the fractal structure, would prevent star formation.

However, galaxy evolution is a dynamical process. \htwo{} clouds can possibly join the visible disk and
contribute to star formation by loosing angular momentum (see Pfenniger 1997 for more details).

\paragraph{}
\textbf{M. F\"{o}rsth} asked: What is the source of the X-ray emission from galaxy clusters?
\paragraph{}
\textbf{A. Tappe} said: Galaxy clusters are among the brightest X-ray sources in the sky. This radiation
comes from very hot gas occupying the space between the galaxies of the cluster. The temperature of the gas
ranges from 20 to 100 million K, which is probably due to heating caused by the interaction with the
galaxies moving through that gas. Its origin is not exactly known. It could in principal be accreted from the
outside or blown out from the galaxies in the cluster.

\paragraph{}
\textbf{Dr. Liu} asked: You mentioned the velocity distribution of galaxies in clusters and the rotation
curves of galaxies as evidence for dark matter. Where is the majority of dark matter? Is it inside the galaxies or
in between them?
\paragraph{}
\textbf{A. Tappe} answered: The rotation curves of disc galaxies suggest that we see only about one tenth of
the gravitating mass. That means that the amount of dark matter is up to 90~\%. Much less is known about the
dark matter fraction in elliptical and dwarf galaxies. Observing galaxy clusters in the visible and X-ray
light, we sense roughly 20 to 30~\% of its total gravitating mass. So, the dark matter fraction is 70 to
80~\%. The hot intergalactic gas giving rise to the X-ray emission of the cluster makes up approximately
10~\% of the gravitating mass.

In general, a ratio of dark to visible matter around 10:1 could still be caused entirely by baryonic dark
matter (consistent with the total amount of visible $\mathrm{\Omega_{vis}\approx0.005}$ and baryonic matter
$\mathrm{\Omega_{b}\approx0.05}$). But there is probably a much larger amount of non-baryonic dark matter
(note that $\mathrm{\Omega_{CDM}\approx0.2}$). Many candidates have been suggested but so far we only know
that these particles are very weakly interacting with themselves and other matter. As a consequence they are
probably distributed rather homogeneously.

Since galaxy clusters contain many galaxy types (spirals, ellipticals, dwarfs, \ldots) in varying amounts,
it is difficult to figure out where the majority of dark matter might be. It depends on how much dark matter
is contained in all these types. By knowing that we would be able to constrain the amount of dark matter
located in between the galaxies. This stuff could be baryonic as well as non-baryonic.

Note finally, that it is probably not easy to distinguish between the ``inside'' and ``outside'' of a galaxy.
The halos and disks can be quite extended, so that sometimes neighbour galaxies are actually overlapping.
\\
\\
\pagebreak

\noindent
{\Large \textbf{Acknowledgements}}
\\
\\
\indent
I am grateful to A.~B. Romeo, J.~H.~Black, T.~Wiklind, L.~Humphreys and B.~T.~Draine for helpful comments.

\end{document}